\documentclass[numreferences]{kluwer}

\newdisplay{guess}{Conjecture}

\begin{document}

\begin{article}

\begin{opening}

\title{Integrable Mappings Related to the Extended Discrete KP Hierarchy\thanks{This
work was supported by INTAS grant 2000-15 and RFBR grant
03-01-00102.}}

\author{ANDREI K. \surname{SVININ}}

\runningauthor{ANDREI K. SVININ}

\runningtitle{Integrable Mappings}

\institute{Institute of System Dynamics and Control Theory, Siberian
Branch of Russian Academy of Sciences, P.O. Box 1233, 664033 Irkutsk,
Russia}

\date{}

\begin{abstract}
We investigate self-similar solutions of the extended discrete KP
hierarchy. It is shown that corresponding ansatzes lead to purely
discrete equations with dependence on some number of parameters
together with equations governing deformations with respect to these
parameters. Some examples are provided. In particular it is shown
that well known discrete first Painlev\'e equation ($\mathrm{dP}_I$)
and its hierarchy arises as self-similar reduction of  Volterra
lattice hierarchy which in turn can be treated as reduction of the
extended discrete KP hierarchy. It is written down equations which
naturally generalize $\mathrm{dP}_I$. It is shown that these discrete
systems describe B\"acklund transformations of Noumi-Yamada systems
of type $A_{2(n-1)}^{(1)}$. We also consider Miura
transformations relating different infinite- and finite-field
integrable mappings. Simplest example of this kind of Miura
transformations is given.
\end{abstract}
\keywords{extended discrete KP hierarchy, self-similar solutions,
discrete Painlev\'e equations}
\end{opening}
\begin{abstract}
We investigate self-similar solutions of the extended discrete KP
hierarchy. It is shown that corresponding ansatzes lead to purely
discrete equations with dependence on some number of parameters
together with equations governing deformations with respect to these
parameters. Some examples are provided. In particular it is shown
that well known discrete first Painlev\'e equation ($\mathrm{dP}_I$)
and its hierarchy arises as self-similar reduction of  Volterra
lattice hierarchy which in turn can be treated as reduction of the
extended discrete KP hierarchy. It is written down equations which
naturally generalize $\mathrm{dP}_I$ in the sense that they have the
$\mathrm{P}_I$ in continuous limit. We also consider Miura
transformations relating different infinite- and finite-field
integrable mappings. Simplest example of this kind of Miura
transformations is given.
\end{abstract}

\section{Introduction}

In Refs. \cite{sv1}, \cite{sv5} we exhibit and investigate
two-parameter class of invariant submanifolds of Darboux-KP (DKP)
chain which is in fact the chain of copies of (differential) KP
hierarchy glued together by Darboux map. The DKP chain is formulated
in terms of formal Laurent series $\{h(i) = z +
\sum_{k=2}^{\infty}h_k(i)z^{-k+1},\;\; a(i) = z +
\sum_{k=1}^{\infty}a_k(i)z^{-k+1} : i\in\mathbb{Z}\}$  and reads as
\cite{mpz}
\begin{equation}
\begin{array}{l}
\partial_p h(i) = \partial H^{(p)}(i),  \\[0.3cm]
\partial_p a(i) = a(i)(H^{(p)}(i+1) - H^{(p)}(i))
\end{array}
\label{DKP}
\end{equation}
with $\{H^{(p)}(i) : p\geq 1\}$ (with fixed value of $i$) being the
currents calculated at the point $h(i)$ \cite{ca2}. Restriction on
invariant submanifold $\mathcal{S}_l^n$ is defined by condition $
z^{l-n+1}a^{[n]}(i)\in{\cal H}_{+}(i),\;\; \forall i\in\mathbb{Z}$
where ${\cal H}_{+}(i) = <1, H^{(1)}(i), H^{(2)}(i),...>$ and
$a^{[k]}(i)$'s are Fa\`a di Bruno discrete iterates defined by
recurrence relation $a^{[k+1]}(i) = a(i)a^{[k]}(i)$ with
$a^{[0]}(i)\equiv 1$. As was shown in Refs. \cite{sv1}, \cite{sv5},
the restriction of DKP chain on $\mathcal{S}_0^n$ leads to linear
discrete systems (see also \cite{sv2})
\begin{equation}
Q_{(n)}^r\Psi = z^r\Psi
\label{aux1}
\end{equation}
and
\begin{equation}
z^{p(n-1)}\partial_p^{(n)}\Psi = (Q_{(n)}^{pn})_{+}\Psi,
\label{aux2}
\end{equation}
with the pair of discrete operators
\[
Q_{(n)}^r \equiv\Lambda^{r} +
\sum_{k\geq 1}q_k^{(n,r)}z^{k(n-1)}\Lambda^{r-kn}
\]
and
\[
(Q_{(n)}^{pn})_{+} \equiv\Lambda^{pn} +
\sum_{k=1}^pq_k^{(n,pn)}z^{k(n-1)}\Lambda^{(p-k)n}
\]
whose coefficients $q_k^{(n,r)} = q_k^{(n,r)}(i)$ uniquely defined as
polynomials in coordinates $a_k(i)$ with the help of the relation
\[
z^r=a^{[r]}(i)+z^{n-1}q_1^{(n,r)}(i)a^{[r-n]}(i) +
z^{2(n-1)}q_2^{(n,r)}(i)a^{[r-2n]}(i) + ...,\;\;
r\in\mathbb{Z}.
\]

The consistency condition of (\ref{aux1}) and (\ref{aux2}) reads as
Lax equation
\begin{equation}
z^{p(n-1)}\partial_p^{(n)}Q_{(n)}^r = [(Q_{(n)}^{pn})_{+}, Q_{(n)}^r]
\label{edkp}
\end{equation}
and can be rewritten in explicit form
\begin{eqnarray}
\partial_p^{(n)}q_k^{(n, r)}(i)&=&Q_{k,p}^{(n,r)}(i)=q_{k+p}^{(n, r)}(i+pn)-
q_{k+p}^{(n, r)}(i)
\nonumber \\[0.3cm]
&&+\sum_{s=1}^{p}q_s^{(n, pn)}(i)\cdot q_{k-s+p}^{(n,r)}(i+(p-s)n)
\nonumber \\[0.3cm]
&&-\sum_{s=1}^{p}q_s^{(n, pn)}(i+r-(k-s+p)n)\cdot
q_{k-s+p}^{(n,r)}(i). 
\label{expl.form}
\end{eqnarray}
Lax equations (\ref{edkp}) naturally contains, for $n=1$, equations
of the discrete KP hierarchy and we call them the extended discrete KP 
hierarchy
\cite{sv2}, \cite{sv1}, \cite{sv5}.

The formula (\ref{expl.form}) is proved to be a container for many
differential-difference systems \cite{sv5} (see, also \cite{sv4}, \cite{sv3}) 
but to derive them from (\ref{expl.form}) one needs to take into account
purely algebraic relations
\[
q_k^{(n,r_1+r_2)}(i)=q_k^{(n,r_1)}(i) +
\sum_{s=1}^{k-1}q_s^{(n,r_1)}(i)\cdot q_{k-s}^{(n,r_2)}(i+r_1-sn) +
q_k^{(n,r_2)}(i+r_1)
\]
\begin{equation}
=q_k^{(n,r_2)}(i) + \sum_{s=1}^{k-1}q_s^{(n,r_2)}(i)\cdot
q_{k-s}^{(n,r_1)}(i+r_2-sn) + q_k^{(n,r_1)}(i+r_2) \label{r}
\end{equation}
which are coded in permutability operator relation
\[
Q_{(n)}^{r_1+r_2} = Q_{(n)}^{r_1}Q_{(n)}^{r_2} =
Q_{(n)}^{r_2}Q_{(n)}^{r_1}.
\]

It is quite obvious that equations (\ref{expl.form}) admit reductions
with the help of simple conditions
\[
q_k^{(n,r)}(i)\equiv 0,\;\; k\geq l+1,\;\; l\geq 1.
\]
which are also consistent with the algebraic relations (\ref{r}).

As was shown in \cite{sv1}, \cite{sv5}, this reductions can be
properly described in geometric setting as double intersections of
invariant manifolds of DKP chain: $\mathcal{S}_{n,r,l} =
\mathcal{S}_0^n\cap\mathcal{S}_{l-1}^{ln-r}$. For example, the
restriction of the DKP chain on $\mathcal{S}_{2,1,1}$ yields Volterra
lattice hierarchy.

Our principal goal in this Letter is to investigate self-similar
solutions of equations (\ref{expl.form}) supplemented by (\ref{r}).
In Section 2, we perform auxiliary calculations aiming to select some
quantities which do not depend on evolution parameters if solution
$\{q_k^{(n,r)}(i)\}$ of the extended discrete KP hierarchy, or
more exactly its subhierarchy corresponding to multi-time
$t^{(n)}$, is invariant with respect to suitable
scaling transformation.
In Section 3, we show that self-similar ansatzes yield a large class of
purely discrete systems supplemented by equations governing
deformations of parameters entering these systems. 
In Section 4, we consider the simplest example corresponding to
Volterra lattice hierarchy and deduce $\mathrm{dP}_I$ and some discrete
equations which can be
treated as higher members in $\mathrm{dP}_I$ hierarchy. We recover there
known relationship between $\mathrm{dP}_I$ and fourth Painlev\'e equation
($\mathrm{P}_{IV}$).
In Section 5, we show one-component discrete equations naturally 
generalizing $\mathrm{dP}_I$
in a sense that they govern B\"acklund transformation for higher order
generalizations of $\mathrm{P}_{IV}$.  
Section 6 is devoted to constructing some class of 
Miura transformations relating
different discrete equations. We provide the reader by simple examples of
such transformations.

\section{Auxiliary calculations}

To prepare a ground for further investigations, we need in following
technical statement

{\it Proposition 2.1.}
By virtue of relations (\ref{r}) with $r_1=pn$, $r_2=sn$ and $k=p+s$,
the relation
\begin{equation}
Q_{p,s}^{(n,pn)}(i)=Q_{s,p}^{(n,sn)}(i). \label{r1}
\end{equation}
with arbitrary $p, s\in\mathbb{N}$ is identity.

{\it Proof.} By direct calculations.

The formula (\ref{r1}), by definition of $Q_{k,p}^{(n,r)}$, can be
rewritten in the form  
\begin{equation}
\partial_s^{(n)}q_p^{(n,pn)}(i)=\partial_p^{(n)}q_s^{(n,sn)}(i),\;\;
\forall p, s\in\mathbb{N}.
\label{from}
\end{equation}
In particular, we have
\begin{eqnarray}
\partial^{(n)}q_p^{(n,pn)}(i)&=&\partial_p^{(n)}q_1^{(n,n)}(i)
=\partial_p^{(n)}\left(\sum_{s=1}^nq_1^{(n,1)}(i+s-1)\right)
\nonumber \\[0.3cm]
&=&\sum_{s=1}^nQ_{1,p}^{(n,1)}(i+s-1). \nonumber
\end{eqnarray}
Each $Q_{1,p}^{(n,1)}(i)$ is uniquely expressed as a polynomial of
$q_k(i)\equiv q_k^{(n,1)}(i)$. So, we can suppose that there exists
the polynomials $\xi_p^{(n)}(i)$ in $q_k(i)$ such that
\begin{equation}
q_p^{(n,pn)}(i)=\sum_{s=1}^n\xi_p^{(n)}(i+s-1),\;\; \forall p\geq 1
\label{that}
\end{equation}
and consequently
\begin{equation}
\partial^{(n)}\xi_p^{(n)}(i)=\partial_p^{(n)}q_1(i)
= Q_{1,p}^{(n,1)}(i). \label{ksi2}
\end{equation}
For example, we can write down the following:
\[
\xi_1^{(n)}(i) = q_1(i),\;\;\; \xi_2^{(n)}(i) = q_2(i)+q_2(i+n) +
q_1(i)\cdot\sum_{s=1-n}^{n-1}q_1(i+s).
\]
Unfortunately we cannot by now to write down the polynomials
$\xi_3^{(n)}(i)$, $\xi_4^{(n)}(i),...$ in explicit form for arbitrary
$n$, but we believe that one can do it for arbitrary concrete values
of $n$. From (\ref{from}) we have the following

{\it Proposition 2.2.} By virtue of equations of motion of the extended
discrete KP hierarchy,
\[
\partial_s^{(n)}\xi_p^{(n)}(i)=\partial_p^{(n)}\xi_s^{(n)}(i),\;\;
\forall s, p\geq 1.
\]

Fixing any integer $p\geq 2$, let us define
\[
\alpha_i^{(n)} = \sum_{k=1}^pkt_k^{(n)}\xi_k^{(n)}(i).
\]
Taking into account the above conjecture, we obtain
\begin{eqnarray}
\partial_s^{(n)}\alpha_i^{(n)}&=&s\xi_s^{(n)}(i)+
\sum_{k=1}^pkt_k^{(n)}\partial_s^{(n)}\xi_k^{(n)}(i)
\nonumber \\[0.3cm]
&=&s\xi_s^{(n)}(i) +\sum_{k=1}^pkt_k^{(n)}\partial_k^{(n)}\xi_s^{(n)}(i)
\nonumber \\[0.3cm]
&=&\left\{s+\sum_{k=1}^pkt_k^{(n)}\partial_k^{(n)}\right\}\xi_s^{(n)}(i),\;\;
s=1,..., p. \label{ai}
\end{eqnarray}

In what follows, we are going to investigate some class of
self-similar solutions of extended discrete KP hierarchy and the
latter auxiliary formula is of great importance from the point of view of
constructing purely discrete systems to which lead corresponding
ansatzes.

\section{Self-similar solutions and integrable mappings}

It is evident, that linear systems  (\ref{aux1}) and (\ref{aux2})
and its consistency relations (\ref{expl.form}) and (\ref{r})
are invariant under group of dilations
\[
q_k^{(n,r)}(i)\rightarrow\epsilon^k q_k^{(n,r)}(i),\;\;
t_l\rightarrow\epsilon^{-l}t_l,\;\;
z\rightarrow\epsilon z,\;\;
\Psi_i\rightarrow\epsilon^i\Psi_i.
\]
In what follows we consider dependence only on finite number of evolution
parameters $t_1^{(n)},..., t_p^{(n)}$. Invariants of this group are
\[
T_l = \frac{t_l^{(n)}}{(pt_p^{(n)})^{l/p}},\;\; l = 1,..., p-1,\;\;
\xi = (pt_p^{(n)})^{1/p}z,
\]
\[
\psi_i = z^i\Psi_i,\;\; x_k^{(n,r)}(i) =
(pt_p^{(n)})^{k/p}q_k^{(n,r)}(i).
\]
From this one gets the ansatzes for self-similar solutions
\begin{equation}
q_k^{(n,r)}(i) = \frac{1}{(pt_p^{(n)})^{k/p}}x_k^{(n,r)}(i),\;\;
\Psi_i = z^i\psi_i(\xi; T_1,..., T_{p-1}).
\label{substitution}
\end{equation}
Here $x_k^{(n,r)}(i)$'s are supposed to be unknown functions of
$T_1,..., T_{p-1}$.
Direct substitution of (\ref{substitution}) into (\ref{expl.form}) gives
\begin{equation}
\partial_{T_l}x_k^{(n,r)}(i) = X_{k,l}^{(n,r)}(i),\;\;
l =1,..., p-1
\label{ev}
\end{equation}
and
\[
Y_{k,p}^{(n,r)}(i) = kx_k^{(n,r)}(i) + T_1X_{k,1}^{(n,r)}(i) +
2T_2X_{k,2}^{(n,r)}(i) + ...
\]
\begin{equation}
+ (p-1)T_{p-1}X_{k,p-1}^{(n,r)}(i)
+ X_{k,p}^{(n,r)}(i) = 0,\;\;
l =1,..., p-1.
\label{ev1}
\end{equation}
Here $X^{(n,r)}_{k,l}(i)$'s are rhs's of (\ref{expl.form}) where
$q^{(n,r)}_k(i)$'s are replaced by  $x^{(n,r)}_k(i)$'s. Moreover, we
must take into account algebraic relations
\[
x_k^{(n,r_1+r_2)}(i)=x_k^{(n,r_1)}(i) +
\sum_{s=1}^{k-1}x_s^{(n,r_1)}(i)\cdot x_{k-s}^{(n,r_2)}(i+r_1-sn) +
x_k^{(n,r_2)}(i+r_1)
\]
\begin{equation}
=x_k^{(n,r_2)}(i) + \sum_{s=1}^{k-1}x_s^{(n,r_2)}(i)\cdot
x_{k-s}^{(n,r_1)}(i+r_2-sn) + x_k^{(n,r_1)}(i+r_2) \label{r11}
\end{equation}
which follows from (\ref{r}).
Corresponding auxiliary linear equations are transformed
to the following form:
\[
\partial_{T_l}\psi = (X_{(n)}^{ln})_{+}\psi,\;\;
l = 1,..., p-1,
\]
\[
\xi\psi_{\xi}=\left\{T_1(X_{(n)}^n)_{+} + 2T_2(X_{(n)}^{2n})_{+}+...\right.
\]
\[
\left.+(p-1)T_{p-1}(X_{(n)}^{(p-1)n})_{+}+(X_{(n)}^{pn})_{+}\right\}\psi,
\]
\[
X_{(n)}^r\psi = \xi\psi,
\]
where
\[
X_{(n)}^r\equiv\xi\Lambda^r + \sum_{k\geq
1}\xi^{1-k}x_k^{(n,r)}\Lambda^{r-kn},\;\; r\in\mathbb{Z}.
\]

Let us now observe that if $\{q_k^{(n,r)}(i)\}$ represent
self-similar solution than by virtue of (\ref{ai}) the quantities
$\alpha_i^{(n)}$ do not depend on $t_1^{(n)},..., t_p^{(n)}$.
Moreover, one can write
\[
\alpha_i^{(n)}(i) = \sum_{k=1}^{p-1}kT_k\zeta_k^{(n)}(i) +
\zeta_p^{(n)}(i),
\]
with $\zeta_k^{(n)}(i) = (pt_p^{(n)})^{k/n}\xi_k^{(n)}(i)$.

So, one can state that $\alpha_i^{(n)}(i)$ in these circumstances do
not depend on parameters $T_1,..., T_{p-1}$.

Simplest situation in which one can easily to derive pure discrete
equations from (\ref{ev1}) supplemented by deformation equations in a
nice form is when $r=1$ and $p=2$. Equations (\ref{ev1}) are
specified in this case as\footnote{In what follows, $x_k(i)\equiv
x_k^{(n,1)}(i)$}
\begin{equation}
Y_{k,2}^{(n,1)}(i) = kx_k(i) + T_1X_{k,1}^{(n,1)}(i) +
X_{k,2}^{(n,1)}(i) = 0. \label{ev2}
\end{equation}
In addition, one must take into account deformation equations
\begin{eqnarray}
\partial_{T_1}x_k(i)&=&X_{k,1}^{(n,1)}(i) = x_{k+1}(i+n) - x_{k+1}(i)
\nonumber \\[0.3cm]
&&+x_k(i)\left( \sum_{s=1}^nx_1(i+s-1) - \sum_{s=1}^nx_1(i+s-kn)
\right). \label{mmm}
\end{eqnarray}

{\it Proposition 3.1.} The relations (\ref{ev2}) read as
infinite-field mapping
\begin{equation}
\alpha_i^{(n)} = T_1x_1(i) + x_1(i)\cdot\sum_{s=1-n}^{n-1}x_1(i+s)
+ x_2(i) + x_2(i+n),
\label{-1}
\end{equation}
\begin{eqnarray}
X_k(i)&=&x_k(i)\left(k + \sum_{s=1}^n\alpha_{i+s-1}^{(n)} -
\sum_{s=1}^n\alpha_{i+s-kn}^{(n)}\right)
\nonumber \\[0.3cm]
&&+x_{k+1}(i+n)\left(T_1+\sum_{s=1}^{2n}x_1(i+s-1)\right)
\nonumber \\[0.3cm]
&&-x_{k+1}(i)\left(T_1 + \sum_{s=1}^{2n}x_1(i+s-(k+1)n)\right)
\nonumber \\[0.3cm]
&&+x_{k+2}(i+2n)-x_{k+2}(i) = 0,\;\; 
k\geq 1 
\label{0}
\end{eqnarray}
with $\alpha_i^{(n)}$'s being arbitrary constants.

{\it Proof.} The proposition is proved by
straightforward calculations. In what follows we use the formula
(\ref{that}) with $p=1, 2$, where $q_k(i)$'s are replaced by
$x_k(i)$'s. We have
\[
X_{k,1}^{(n,1)}(i)=x_{k+1}(i+n) - x_{k+1}(i) +
x_k(i)\left(x_1^{(n,n)}(i) - x_1^{(n,n)}(i+1-kn)\right)
\]
\begin{eqnarray}
&=&x_{k+1}(i+n) - x_{k+1}(i) +
x_k(i)\left(\sum_{s=1}^n\xi_1^{(n)}(i+s-1) \right. \nonumber \\[0.3cm]
&&-\left. \sum_{s=1} ^n\xi_1^{(n)}(i+s-kn)\right), \label{1}
\end{eqnarray}
\[
X_{k,2}^{(n,1)}(i)=x_{k+2}(i+2n) + x_1^{(n,2n)}(i)x_{k+1}(i+n) +
x_2^{(n,2n)}(i)x_k(i)
\]
\[-x_{k+2}(i) - x_1^{(n,2n)}(i+1-(k+1)n)x_{k+1}(i) -
x_2^{(n,2n)}(i+1-kn)x_k(i)
\]
\[
=x_{k+2}(i+2n) + x_{k+1}(i+n)\cdot\sum_{s=1}^{2n}x_1(i+s-1) +
x_k(i)\cdot\sum_{s=1}^n\xi_2^{(n)}(i+s-1)
\]
\begin{equation}
-x_{k+2}(i) - x_{k+1}(i)\cdot\sum_{s=1}^{2n}x_1(i+s-(k+1)n) -
x_k(i)\cdot\sum_{s=1}^n\xi_2^{(n)}(i+s-kn). \label{2}
\end{equation}
One can easily check that substituting (\ref{1}) and (\ref{2}) into
(\ref{ev2}) gives (\ref{0}) with $\alpha_i^{(n)}$ given by
(\ref{-1}). Moreover, it is already proved that $\alpha_i^{(n)}$'s do
not depend on $t_1^{(n)},..., t_p^{(n)}$ and therefore on $T_1,...,
T_{p-1}$.

It is obvious that the system (\ref{-1}) and (\ref{0}) admits as well as
equations (\ref{mmm}) reduction with the help of simple condition
\[
x_k(i)\equiv 0,\;\; k>l,\;\; l\geq 1.
\]
Then $l$-th equation in (\ref{0}) is specified as
\[
x_l(i)\cdot\left\{
l + \sum_{s=1}^n\alpha_{i+s-1}^{(n)} -
\sum_{s=1}^n\alpha_{i+s-ln}^{(n)}
\right\} = 0.
\]
Since it is supposed that $x_l(i)\not\equiv 0$ then the constants
$\alpha_i^{(n)}$ are forced to be subjects of constraint
\[ l +
\sum_{s=1}^n\alpha_{i+s-1}^{(n)} - \sum_{s=1}^n\alpha_{i+s-ln}^{(n)}
= 0.
\]

\section{$\mathrm{dP}_I$ and its hierarchy}

Let us consider, as a simplest example, the case corresponding to
Volterra lattice hierarchy, that is $n=2, r=1, l=1$. Take $p=2$. The
equations (\ref{ev2}) and (\ref{mmm}) are written down as follows:
\begin{equation}
x_i^{\prime} = x_i(x_{i+1} - x_{i-1}),\;\;
\phantom{q}^{\prime}\equiv\partial/\partial T_1,
\label{evr2}
\end{equation}
\begin{equation}
x_i + T_1x_i\left\{x_1^{(2,2)}(i) - x_1^{(2,2)}(i-1)\right\} +
x_i\left\{x_2^{(2,4)}(i) - x_2^{(2,4)}(i-1)\right\} = 0.
\label{ev5}
\end{equation}
Here we denote $x_i = x_1^{(2,1)}(i)$. Using (\ref{r1}) one calculates
\[
x_1^{(2,2)}(i) = x_i + x_{i+1},
\]
\[
x_2^{(2,4)}(i) = x_i(x_{i-1} + x_i + x_{i+1})
+ x_{i+1}(x_i + x_{i+1} + x_{i+2}).
\]
Taking into account these relations, the equation (\ref{ev5}) turns
into
\begin{equation}
\begin{array}{l}
T_1x_i + x_i(x_{i-1} + x_i + x_{i+1}) = \alpha_i^{(2)}, \\[0.3cm]
1+\alpha^{(2)}_{i+1}-\alpha^{(2)}_{i-1} = 0.
\end{array}
\label{dp1}
\end{equation}
One can rewrite (\ref{dp1}) as
\begin{equation}
x_{i-1} + x_i + x_{i+1} = -T_1 + \frac{\alpha^{(2)}_i}{x_i},
\label{dp2}
\end{equation}
where $\alpha_i^{(2)}$'s are some constants forced to be subjects of
the quasi-periodicity constraint: $\alpha^{(2)}_{i+2} =
\alpha^{(2)}_i - 1$. One can immediately  to write down the solution
of this equation: $\alpha^{(2)}_i = \alpha - \frac{1}{2}i +
\beta(-1)^i$, where $\alpha$ and $\beta$ are some complex constants. Provided
these conditions, (\ref{dp2}) is $\mathrm{dP}_I$ \cite{gr1}.

Observe that evolution equation (\ref{evr2}) with (\ref{dp2}) turns into
\[
x_i^{\prime} = 2x_ix_{i+1} + x_i^2 + T_1x_i - \alpha^{(2)}_i.
\]
It can be easily checked that together with (\ref{dp2}) this lattice
is equivalent to the pair of ordinary first-order differential
equations
\begin{equation}
\begin{array}{l}
w_1^{\prime} = 2w_1w_2 + w_1^2 + T_1w_1 + a, \\[0.3cm]
w_2^{\prime} = -2w_1w_2 - w_2^2 - T_1w_2 - b
\end{array}
\label{twin}
\end{equation}
with discrete symmetry transformation (cf. \cite{gr2})
\[
\overline{w}_1 = w_2,\;\;
\overline{w}_2 = - w_1 - w_2 - T_1 -\frac{b}{w_2},\;\;
\overline{a} = b,\;\;
\overline{b} = a + 1,
\]
where $w_1\equiv x_i,\; w_2 = x_{i+1},\; a\equiv -\alpha^{(2)}_i,\;
b\equiv -\alpha^{(2)}_{i+1}$ for some fixed (but arbitrary) value
$i=i_0$. In turn  the system (\ref{twin}) is equivalent to
second-order equation
\[
w^{\prime\prime} = \frac{(w^{\prime})^2}{2w} + \frac{3}{2}w^3 + 2T_1w^2 +
\left(\frac{T_1^2}{2} + a - 2b + 1\right)w - \frac{a^2}{2w},\;\;
w\equiv w_1
\]
with corresponding symmetry transformation
\[
\overline{w} =
\frac{w^{\prime} - w^2 - T_1w - a}{2w},\;\;
\overline{a} = b,\;\;
\overline{b} = a + 1.
\]
In fact this is $\mathrm{P}_{IV}$ with B\"acklund transformation
\cite{gr2}. By rescaling  $T_1\rightarrow\sqrt{2}T_1,\; w\rightarrow
w/\sqrt{2}$ it can be turned to following canonical form:
\begin{equation}
w^{\prime\prime} = \frac{(w^{\prime})^2}{2w} + \frac{3}{2}w^3 + 4T_1w^2 +
2(T_1^2 + a - 2b + 1)w - \frac{2a^2}{w}
\label{P42}
\end{equation}
\[
\overline{w} =
\frac{w^{\prime} - w^2 - 2T_1w - 2a}{2w},\;\;
\overline{a} = b,\;\;
\overline{b} = a + 1.
\]
As a result, we obtain the well-known relationship between
$\mathrm{dP}_I$ (\ref{dp2}) and $\mathrm{P}_{IV}$ (\ref{P42}) (see,
for example \cite{gr3}).

{\it Remark 4.1.} The equations (\ref{twin}) can be interpreted as
self-similar reduction of Levi system \cite{marikhin}
\[
\begin{array}{l}
v_{1t_2} = (-v_1^{\prime} + v_1^2 + 2v_1v_2)^{\prime}, \\[0.3cm]
v_{2t_2} = (v_2^{\prime} + v_2^2 + 2v_1v_2)^{\prime}
\end{array}
\]
with the help of the ansatz
\[
v_k=\frac{1}{(2t_2)^{1/2}}w_k(T_1),\;\;
k = 1, 2.
\]

Let us discuss now the higher members in $\mathrm{dP}_I$ hierarchy.
To construct them, one needs to consider the cases $p = 3, 4$ and so
on, that is
\[
x_i + T_1x_i\left\{x_1^{(2,2)}(i) - x_1^{(2,2)}(i-1)\right\}  +
2T_2x_i\left\{x_2^{(2,4)}(i) - x_2^{(2,4)}(i-1)\right\}
\]
\begin{equation}
+ x_i\left\{x_3^{(2,6)}(i) - x_2^{(2,6)}(i-1)\right\}= 0, \label{1t}
\end{equation}
\[
x_i + T_1x_i\left\{x_1^{(2,2)}(i) - x_1^{(2,2)}(i-1)\right\} +
2T_2x_i\left\{x_2^{(2,4)}(i) - x_2^{(2,4)}(i-1)\right\}
\]
\begin{equation}
+ 3T_3x_i\left\{x_3^{(2,6)}(i) - x_2^{(2,6)}(i-1)\right\} +
\left\{x_4^{(2,8)}(i) - x_4^{(2,8)}(i-1)\right\}= 0. \label{2t}
\end{equation}
We have calculated, to wit the following
\[
x_3^{(2,6)}(i) = \zeta_3^{(2)}(i) + \zeta_3^{(2)}(i+1)
\]
with
\[
\zeta_3^{(2)}(i) = x_i\left\{\zeta_2^{(2)}(i-1) + \zeta_2^{(2)}(i) +
\zeta_2^{(2)}(i+1) + x_{i-1}x_{i+1} \right\}
\]
and
\[
x_4^{(2,8)}(i) = \zeta_4^{(2)}(i) + \zeta_4^{(2)}(i+1)
\]
with
\[
\zeta_4^{(2)}(i) = x_i\left\{\zeta_3^{(2)}(i-1)\right.
\]
\[
\left. + \zeta_3^{(2)}(i) + \zeta_3^{(2)}(i+1) +
x_{i-1}x_{i+1}(x_{i-2} + x_{i-1} + x_{i} + x_{i+1} +x_{i+2})
\right\}.
\]
Substituting the latter formulas into (\ref{1t}) and (\ref{2t}) we
get the higher members in $\mathrm{dP}_I$ hierarchy in the form
\[
T_1x_i + 2T_2\zeta_2^{(2)}(i) + \zeta_3^{(2)}(i) = \alpha_i^{(2)},
\]
\[
T_1x_i + 2T_2\zeta_2^{(2)}(i) + 3T_3\zeta_3^{(2)}(i) +
\zeta_4^{(2)}(i) = \alpha_i^{(2)},
\]
with corresponding constants $\alpha_i^{(2)}$ being subjected to the
constraint $1 + \alpha_{i+1}^{(2)} - \alpha_{i-1}^{(2)} = 0$. These
results entirely correspond to that of the work \cite{cr}.

\section{Generalizations of $\mathrm{dP}_I$}

Let us consider one-field reductions of the system given by equations
(\ref{-1}) and (\ref{0}) when $n$ is arbitrary. The analogues of the
equations (\ref{evr2}) and (\ref{ev5}) in this case are
\begin{equation}
x_i^{\prime} = x_i\left(\sum_{s=1}^{n-1}x_{i+s} -
\sum_{s=1}^{n-1}x_{i-s}\right),
\label{bv2}
\end{equation}
\[
x_i + T_1x_i\left\{x_1^{(n,n)}(i) - x_1^{(n,n)}(i+1-n)\right\}
\]
\begin{equation}
+ x_i\left\{x_2^{(n,2n)}(i) - x_2^{(n,2n)}(i+1-n)\right\} = 0.
\label{v3}
\end{equation}
We take into account that
\[
x_1^{(n,n)}(i) = \sum_{s=1}^n x_{i+s-1},\;\;
x_2^{(n,2n)}(i) = \sum_{s=1}^n x_{i+s-1}\left(
\sum_{s_1=1}^{2n-1}x_{i+s_1+s-n-1}\right).
\]
Then the equation (\ref{v3}) can be cast into the form
\[
T_1x_i + x_i(x_{i+1-n} + ... +  x_{i+n-1}) = \alpha^{(n)}_i,
\]
\begin{equation}
1 + \sum_{s=1}^{n-1}\alpha^{(n)}_{i+s} -
\sum_{s=1}^{n-1}\alpha^{(n)}_{i-s} = 0. \label{4}
\end{equation}
One can write the solution of (\ref{4}) as
\begin{equation}
\alpha^{(n)}_i=\alpha-\frac{1}{(n-1)n}i+\beta \omega^i
\label{ai4}
\end{equation}
with arbitrary constants $\alpha,\;\beta\in\mathbb{C}$ and 
$\omega=\sqrt[n]{1}=\exp(2\pi\sqrt{-1}/n)$.

So, one concludes that in this case self-similar ansatz leads to
equation
\begin{equation}
x_{i+1-n} + ... + x_{i+n-1} = -T_1 + \frac{\alpha^{(n)}_i}{x_i},
\label{dp3}
\end{equation}
where the constants $\alpha^{(n)}_i$'s are given by (\ref{ai4}).

Standard analysis of singularity confinement shows that this property for
(\ref{dp3}) is valid provided that
\begin{equation}
\alpha^{(n)}_{i+n} - \alpha^{(n)}_i = \alpha^{(n)}_{i+2n-1} -
\alpha^{(n)}_{i+n-1}. \label{usl}
\end{equation}
but one can show that this equation do not contradict to (\ref{4}),
but it is more general. Indeed, it follows from (\ref{4}) that
\begin{equation}
-1 = \sum_{s=1}^n\alpha^{(n)}_{i+s+n-1} -
\sum_{s=1}^n\alpha^{(n)}_{i+s} = \sum_{s=1}^n\alpha^{(n)}_{i+s+n-2} -
\sum_{s=1}^n\alpha^{(n)}_{i+s-1}. \label{usl1}
\end{equation}
Second equality in (\ref{usl1}) can be rewritten in the form
\[
\sum_{s=2}^{n-1}\alpha^{(n)}_{i+s+n-1} + \alpha^{(n)}_{i+2n-1} -
\sum_{s=1}^{n-2}\alpha^{(n)}_{i+s} - \alpha^{(n)}_{i+n-1}
\]
\[
= \alpha^{(n)}_{i+n} + \sum_{s=3}^{n}\alpha^{(n)}_{i+s+n-2} -
\alpha^{(n)}_i - \sum_{s=2}^{n-1}\alpha^{(n)}_{i+s-1}.
\]
From the latter one obtains (\ref{usl}).

Finally, let us show that equations (\ref{bv2}), (\ref{4}) and (\ref{dp3})
are equivalent to higher order Painlev\'e equation of type $A_l^{(1)}$ 
\cite{noumi2} with $l=2(n-1)$, i.e. higher order generalization of $\mathrm{P}_{IV}$. 
Following the line of previous section, one can observe that equations
(\ref{bv2}), (\ref{4}) and (\ref{dp3}) are equivalent to the system
\begin{equation}
\begin{array}{l}
\displaystyle
w_k^{\prime}=w_k\left(
2\sum_{s=1}^{n-1}w_{k+s}+w_k+T_1\right)+a_k, \\
\displaystyle
w_{k+n-1}^{\prime}=-w_{k+n-1}\left(
2\sum_{s=1}^{n-1}w_{k+s-1}+w_{k+n-1}+T_1\right)-a_{k+n-1} \\
(k=1,..., n-1)
\end{array}
\label{nonsym}
\end{equation}
supplemented by B\"acklund transformation
\begin{equation}
\begin{array}{l}
\displaystyle
\overline{w}_k=w_{k+1},\;\; k=1,..., 2n-3,\;\; \overline{w}_{2(n-1)}
=-\sum_{s=1}^{2(n-1)}w_s-\frac{a_n}{w_n}-T_1, \\
\displaystyle
\overline{a}_k=a_{k+1},\;\; k=1,..., 2n-3,\;\; 
\overline{a}_{2(n-1)}=\sum_{s=1}^{n-1}a_s-\sum_{s=1}^{n-2}a_{n+s}+1.
\end{array}
\label{bt}
\end{equation}
Here we identify $w_k=x_{i+k-1}$ and $a_k=-\alpha^{(n)}_{i+k-1}$. 
To write down the equations (\ref{nonsym}) in symmetric form one need 
to introduce new variables $\{f_0, f_1,..., f_{2(n-1)}\}$ by
\[
f_{2k}=-w_k,\;\; 
f_{2k-1}=-w_{k+n-1}\;\;
(k=1,..., n-1),\;\;
f_0=\sum_{s=1}^{2(n-1)}w_s+T_1.
\]
It is evident that $\sum_{s=0}^{2(n-1)}f_s=T_1$. By straightforward 
calculations one can check that the system (\ref{nonsym}) cast into following
symmetric form 
\begin{equation}
f_k^{\prime}=
f_k\left(\sum_{r=1}^{n-1}f_{k+2r-1}-\sum_{r=1}^{n-1}f_{k+2r}\right)+c_k\;\;
(k=0,..., 2(n-1))
\label{noumi}
\end{equation}
where subscripts are supposed to be an elements of 
$\mathbb{Z}/(2n-1)\mathbb{Z}$. The constants $c_k$'s are related with $a_k$'s
by the relations
\[
\begin{array}{l}
c_{2k}=-a_k,\;\;
c_{2k-1}=a_{k+n-1},\;\;
k=1,..., n-1, \\
\displaystyle
c_0=\sum_{s=1}^{n-1}a_s-\sum_{s=1}^{n-1}a_{s+n-1}+1,\;\;
\sum_{s=0}^{2(n-1)}c_s=1.
\end{array}
\]

The system  (\ref{noumi}) is nothing but Noumi-Yamada system of $A_l^{(1)}$
type with $l=2(n-1)$. As for B\"acklund transformation (\ref{bt}), 
one can verify that it 
coinsides with an element $T=s_1\pi^2$ of the extended affine Weil group 
$\tilde{W}=<\pi, s_0,..., s_{2(n-1)}>$ \cite{noumi2}. 

For the sake of completness, let us provide the reader by suitable
information on some representation of affine Weil group of type 
$A_l^{(1)}$ which is useful for constructing of B\"acklund transformations
of $\mathrm{P}_{IV}$ and $\mathrm{P}_{V}$ and its generalizations. 
The action of automophisms $\{s_0,..., s_l\}$
on the field of rational functions in variables 
$\{c_0,..., c_l\}$ and $\{f_0,..., f_l\}$ can be defined as follows 
\cite{noumi2}: 
\[
\begin{array}{l}
s_k(c_k)=-c_k,\;\;
s_k(c_l)=c_l+c_k\;\;
(l=k\pm 1),
s_k(c_l)=c_l\;\;
(l\neq k, k\pm 1),  \\
\displaystyle
s_k(f_k)=f_k,\;\;
s_k(f_l)=f_l\pm\frac{c_k}{f_k}\;\;
(l=k\pm 1),
s_k(f_l)=f_l\;\;
(l\neq k, k\pm 1).
\end{array}
\]
One also defines an automorphism $\pi$ by the rules $\pi(c_k)=c_{k+1}$ and
$\pi(f_k)=f_{k+1}$. It is known by \cite{noumi1} that this set of 
automorphisms define a representation of the extended affine Weil group 
$\tilde{W}$ and represents a collection of B\"acklund transformations
of the system (\ref{noumi}).

\section{Miura transformations}

We showed in Ref. \cite{sv5} that symmetry transformation
\[
g_k : \left\{
\begin{array}{l}
a(i)\rightarrow z^{1-k}a^{[k]}(ki) \\[0.3cm]
h(i)\rightarrow h(ki)
\end{array}
\right.
\]
on the space of DKP chain solutions is a suitable basis for
constructing of some class of lattice Miura transformations. 
In the language of
invariant submanifolds we have
$g_k(\mathcal{S}^{kn}_0)\subset\mathcal{S}_0^n$ and
$g_k(\mathcal{S}_{kn,r,l})\subset\mathcal{S}_{n,r,kl}$. In terms of
coordinate $q_s^{(n,r)}(i)$ one can use simple relation
$\overline{q}_s^{(n,r)}(i) = q_s^{(kn,kr)}(ki)$  but to get lattice
Miura transformation in the form $\overline{q}_k^{(n,r)}(i) =
F(\{q_s^{(kn,r)}(ki)\})$ one needs to make use algebraic relations
(\ref{r}). For example, Miura transformation
$g_2(\mathcal{S}^2_0)\subset\mathcal{S}_0^1$ reads
\[
\overline{q}_k^{(1,1)}(i) = q_k^{(2,2)}(2i)|_{(\ref{r})}
\]
\begin{equation}
= q_k^{(2,1)}(2i) + \sum_{s=1}^{k-1}q_s^{(2,1)}(2i)\cdot
q_{k-s}^{(2,1)}(2i-2s+1)+ q_k^{(2,1)}(2i+1).  \label{in}
\end{equation}

We believe that this approach can be applied for constructing Miura
transformations relating different integrable mappings. Let us show
the simplest example. Denote $y_k(i) = x_k^{(1,1)}(i)$ and $\beta_i =
\alpha_i^{(1)}$. The equations (\ref{-1}) and (\ref{0}) defining
integrable mapping in this case are specified as
\begin{equation}
\beta_i = T_1y_1(i) + y_1^2(i) + y_2(i) + y_2(i+1), \label{DP3}
\end{equation}
\begin{eqnarray}
Y_k(i)&=&y_k(i)\left( k + \beta_i - \beta_{i-k+1}\right) +
y_{k+1}(i+1)\left(T_1 + y_1(i) + y_1(i+1)\right) \nonumber \\[0.3cm]
&&-y_{k+1}(i)\left(T_1 + y_1(i-k) + y_1(i-k+1)\right) \nonumber \\[0.3cm]
&&+y_{k+2}(i+2) - y_{k+2}(i) = 0,\;\; k\geq 1. 
\label{DP4}
\end{eqnarray}
Denote $x_k(i) = x_k^{(2,1)}(i)$ and $\alpha_i = \alpha_i^{(2)}$.
Infinite-field discrete system in the case $n=2$ is
\begin{equation}
\alpha_i=T_1x_1(i)+x_1(i)(x_1(i-1)+x_1(i)+x_1(i+1))+x_2(i)
+x_2(i+2), \label{DP1}
\end{equation}
\begin{eqnarray}
X_k(i)&=&x_k(i)\left(k+\alpha_i+\alpha_{i+1}-\alpha_{i-2k+1}-
\alpha_{i-2k+2}\right) 
\nonumber \\[0.3cm]
&&+x_{k+1}(i+2)\left(T_1+x_1(i)+x_1(i+1)+x_1(i+2)+
x_1(i+3)\right) 
\nonumber \\[0.3cm]
&&-x_{k+1}(i)\left(T_1+x_1(i-2k-1)+x_1(i-2k)\right. 
\nonumber \\[0.3cm]
&&+\left.x_1(i-2k+1)+x_1(i-2k+2)\right) 
\nonumber \\[0.3cm]
&&+x_{k+2}(i+4)-x_{k+2}(i) = 0,\;\; 
k\geq 1. 
\label{DP2}
\end{eqnarray}
Restriction of the latter system on $\mathcal{S}_{2,1,1}$ gives
$\mathrm{dP}_I$ (\ref{dp2}), while the restriction of the system
(\ref{DP3}) and (\ref{DP4}) on $\mathcal{S}_{1,1,2}$ yields two-field
system
\[
\beta_i = T_1y_1(i) + y_1^2(i) + y_2(i) + y_2(i+1),
\]
\begin{equation}
y_1(i) + y_2(i+1)(T_1 + y_1(i) + y_1(i+1)) - y_2(i)(T_1 + y_1(i-1) +
y_1(i)) = 0, \label{st}
\end{equation}
\[
2 + \beta_i - \beta_{i-1} = 0.
\]

One can easily write down Miura transformation relating
infinite-field mappings given by pairs of equations 
(\ref{DP3}), (\ref{DP4}) and (\ref{DP1}),
(\ref{DP2}), respectively, by replacing $\overline{q}_1^{(1,1)}(i)\rightarrow
y_k(i)$ and $q_1^{(2,1)}(i)\rightarrow x_k(i)$ in (\ref{in}), to wit
\begin{equation}
y_k(i) = x_k(2i) + \sum_{s=1}^{k-1}x_s(2i)\cdot x_{k-s}(2i-2s+1)+
x_k(2i+1). \label{sub}
\end{equation}
Moreover we have $\beta_i = \alpha_{2i} + \alpha_{2i+1}$. By
straightforward but tedious calculations one can check that
substituting (\ref{sub}) into (\ref{DP4}) gives
\[
Y_k(i)|_{(\ref{sub})} = X_k(2i) + \sum_{s=1}^{k-1}X_s(2i)\cdot
X_{k-s}(2i-2s+1)+ X_k(2i+1) = 0.
\]

As for transformation
$g_2(\mathcal{S}_{2,1,1})\subset\mathcal{S}_{1,1,2}$ relating
$\mathrm{dP}_I$ with the system (\ref{st}), we have it in the form
\[
y_1(i)=x(2i)+x(2i+1),\;\; y_2(i)=x(2i-1)x(2i),\;\; 
\beta_i=\alpha_{2i}+\alpha_{2i+1}.
\]

\section{Concluding remarks}

We showed that in the simplest case corresponding to the restriction of 
dynamics on phase-space of the DKP chain
on invariant submanifold ${\cal S}_{n,1,1}$ one derives one-field discrete dynamical system governing
B\"acklund transformation of Noumi-Yamada system of $A_{2(n-1)}^{(1)}$ type
which is a natural higher-order generalization of $\mathrm{P}_{IV}$. As is known
higher-order
generalizations of $\mathrm{P}_{V}$ correspond to affine Weil groups of type $A_l^{(1)}$ with
odd $l$. 

On the other hand, these generalizations are also described by 
Veselov-Shabat periodic dressing chains
\[
r_i^{\prime}+r_{i+1}^{\prime}=r_i^2-r_{i+1}^2+\alpha_i,\;\;
r_{i+N}=r_i,\;\;
\alpha_{i+N}=\alpha_i
\]
with odd $N\geq 3$ for $\mathrm{P}_{IV}$ and with even $N\geq 4$ for $\mathrm{P}_{V}$, 
respectively \cite{veselov}, \cite{adler}. So, in a sense $\mathrm{P}_{IV}$ and $\mathrm{P}_{V}$ go in 
parallel in these two settings. We can expect that our approach also covers
$\mathrm{P}_{V}$ and its higher-order generalizations, but to save the space we leave 
this question for subsequent publications.

\section*{Acknowledgement}
I am grateful to the anonymous referee for bringing to my attention 
Refs. \cite{noumi1} and \cite{noumi2} and for a number of
remarks which enabled the presentation of the Letter to be improved.

\end{article}

\end{document}